\documentclass[12pt]{iopart}

\usepackage[lmargin=2.5cm,rmargin=2.5cm,tmargin=3cm,bmargin=2cm]{geometry}
\usepackage{amssymb}
\usepackage{graphicx}
\usepackage{graphics}
\usepackage{appendix}
\usepackage{multirow}
\usepackage{hhline}
\usepackage[normalem]{ulem}
\usepackage[usenames, dvipsnames]{color}
\usepackage{relsize}
\usepackage{color}
\usepackage[font=small]{caption}
\usepackage{subcaption}
\usepackage{xspace}
\usepackage{cite}

\usepackage{hyperref}
\hypersetup{
    colorlinks=true,
    linkcolor=blue,
    filecolor=blue,      
    urlcolor=blue,
    citecolor=blue
}

\usepackage[capitalize]{cleveref}
\crefformat{footnote}{#2\footnotemark[#1]#3}

\usepackage{dcolumn}
\newcolumntype{d}[1]{D{.}{.}{#1}}

\newcommand{\jg}[1]{\textcolor{red}{#1}}
\renewcommand{\jg}[1]{#1}
\newcommand{\iaea}[1]{\textcolor{red}{#1}}
\renewcommand{\iaea}[1]{#1}
\newcommand{\yk}[1]{\textcolor{red}{#1}}
\renewcommand{\yk}[1]{#1}

\newcommand{\nfadd}[1]{\textcolor{red}{#1}}
\renewcommand{\nfadd}[1]{#1\unskip}
\newcommand{\nfsub}[1]{\sout{#1}}
\renewcommand{\nfsub}[1]{\unskip}

\newcommand{\Alfven}{Alfv\'{e}n\xspace}
\newcommand{\AlfvenEigenmode}{\Alfven Eigenmode\xspace}
\newcommand{\AlfvenEigenmodes}{\Alfven Eigenmodes\xspace}
\renewcommand{\AE}{AE\xspace}
\newcommand{\AEs}{AEs\xspace}
\newcommand{\TAE}{TAE\xspace}
\newcommand{\TAEs}{TAEs\xspace}

\newcommand{\GAEs}{GAEs\xspace}
\newcommand{\EAE}{EAE\xspace}
\newcommand{\EAEs}{EAEs\xspace}
\newcommand{\AEADiagnostic}{\AlfvenEigenmode Active Diagnostic\xspace}
\newcommand{\AEAD}{AEAD\xspace}

\newcommand{\xpoint}{X-point\xspace}
\renewcommand{\etal}{\emph{et al}\xspace}

\newcommand{\EP}{EP\xspace}
\newcommand{\EPs}{EPs\xspace}
\newcommand{\FI}{FI\xspace}
\newcommand{\FIs}{FIs\xspace}

\newcommand{\Hethree}{^{3}\mathrm{He}}

\newcommand{\EFIT}{EFIT\xspace}

\newcommand{\HELENA}{HELENA\xspace}
\newcommand{\CSCAS}{CSCAS\xspace}

\newcommand{\NOVAK}{{NOVA-K}\xspace}
\newcommand{\NBI}{NBI\xspace}
\newcommand{\DNBI}{(\mathrm{D}_\mathrm{NBI})}
\newcommand{\ICRH}{ICRH\xspace}
\newcommand{\TRANSP}{TRANSP\xspace}
\newcommand{\Hmode}{{H-mode}\xspace}
\newcommand{\Lmode}{{L-mode}\xspace}
\newcommand{\Mmode}{{L-mode}\xspace}

\newcommand{\wo}{\omega_0}
\newcommand{\f}{f}
\newcommand{\fo}{f_0}
\newcommand{\go}{\gamma/\wo}
\newcommand{\g}{\gamma}

\newcommand{\dgo}{\Delta(\gamma/\wo)}
\newcommand{\n}{n}
\newcommand{\m}{m}

\newcommand{\absn}{\vert\n\vert}
\newcommand{\q}{q}
\newcommand{\qo}{q_0}
\newcommand{\qnf}{q_{95}}
\newcommand{\nnf}{n_\mathrm{e95}}
\newcommand{\gradnnf}{\nabla\nnf}
\newcommand{\Pnbi}{P_{\mathrm{\NBI}}}

\newcommand{\Picrh}{P_{\mathrm{\ICRH}}}

\newcommand{\Bo}{B_0}

\renewcommand{\ne}{n_{\mathrm{e}}}
\newcommand{\neo}{n_{\mathrm{e}0}}
\renewcommand{\ni}{n_{\mathrm{i}}}
\newcommand{\Te}{T_{\mathrm{e}}}
\newcommand{\Teo}{T_{\mathrm{e}0}}
\newcommand{\Ti}{T_{\mathrm{i}}}

\newcommand{\Ip}{I_\mathrm{p}}

\newcommand{\snf}{s_{95}}

\newcommand{\vA}{v_\mathrm{A}}

\renewcommand{\t}{t}
\newcommand{\psin}{\psi_\mathrm{N}}
\newcommand{\rw}{r_\mathrm{w}}
\newcommand{\vpar}{v_\parallel}
\newcommand{\vo}{v_0}
\newcommand{\pfi}{p_\mathrm{FI}}
\newcommand{\wfi}{\omega_\mathrm{*FI}}
\newcommand{\nHethree}{n_\mathrm{He3}}
\newcommand{\vte}{v_\mathrm{te}}
\newcommand{\vti}{v_\mathrm{ti}}

\newcommand{\halfwidth}{0.49\columnwidth}

\newcommand{\SI}[2]{#1~\mathrm{#2}}
\newcommand{\rd}{\mathrm{d}}

\newcommand{\abs}[1]{\vert #1 \vert}

\renewcommand{\xi}{x_i}

\begin{document}

    \title[]{A novel measurement of marginal \AlfvenEigenmode stability during high power auxiliary heating in JET}
    

\newcommand{\iPSFC}{$^1$\xspace}
\newcommand{\iCCFE}{$^2$\xspace}
\newcommand{\iEPFL}{$^3$\xspace}
\newcommand{\iESPCI}{$^4$\xspace}
\newcommand{\iUkraine}{$^6$\xspace}
\newcommand{\iCEA}{$^6$\xspace}
\newcommand{\iUCI}{$^a$\xspace}
\newcommand{\iBelgium}{$^7$\xspace}
\newcommand{\iLisbon}{$^8$\xspace}
\newcommand{\iMilan}{$^9$\xspace}
\newcommand{\iIPST}{$^{10}$\xspace}
\newcommand{\iPPPL}{$^5$\xspace}
\newcommand{\iSlovenia}{$^{11}$\xspace}
\newcommand{\iJET}{*\xspace}

\newcommand{\PSFC}{\iPSFC Plasma Science and Fusion Center, Massachusetts Institute of Technology, Cambridge, MA, USA\xspace}

\newcommand{\EPFL}{\iEPFL Ecole Polytechnique F\'{e}d\'{e}rale de Lausanne (EPFL), Swiss Plasma Center (SPC), CH-1015 Lausanne, Switzerland} 

\newcommand{\CCFE}{\iCCFE Culham Centre for Fusion Energy, Culham Science Centre, Abingdon, UK} 

\newcommand{\CEA}{\iCEA CEA, IRFM, F-13108 Saint-Paul-lez-Durance, France}

\newcommand{\UCI}{\iUCI Department of Physics and Astronomy, UC Irvine, Irvine, USA}

\newcommand{\Ukraine}{\iUkraine Institute of Plasma Physics, NSC KIPT, 310108 Kharkov, Ukraine}

\newcommand{\Belgium}{\iBelgium Laboratory for Plasma Physics, LPP-ERM/KMS, TEC Partner, 1000 Brussels, Belgium}

\newcommand{\Milan}{\iMilan Dipartimento di Fisica, Universit\'{a} di Milano-Bicocca, 20126 Milan, Italy}

\newcommand{\IPST}{\iIPST Institute for Plasma Science and Technology, National Research Council, 20125, Milan, Italy}

\newcommand{\ESPCI}{\iESPCI Ecole Sup\'{e}rieure de Physique et de Chimie Industrielles de la Ville de Paris, 75231 Paris Cedex 05, France}

\newcommand{\Lisbon}{\iLisbon Instituto de Plasmas e Fus\~{a}o Nuclear, Instituto Superior T\'{e}cnico, Univ. de Lisboa, Lisbon, Portugal}

\newcommand{\PPPL}{\iPPPL Princeton Plasma Physics Laboratory, Princeton, NJ, USA}

\newcommand{\Slovenia}{\iSlovenia Jo\v{z}ef Stefan Institute, Ljubljana, Slovenia}

\newcommand{\JET}{\iJET See the author list of ``Overview of JET results for optimising ITER operation'' by J.~Mailloux \etal to be published in \emph{Nuclear Fusion} special issue: Overview and Summary Papers from the 28th Fusion Energy Conference (Nice, France, 10-15 May 2021)}

\author{R.A.~Tinguely\iPSFC\footnote{Author to whom correspondence should be addressed: rating@mit.edu}, 
    N.~Fil\iCCFE, 
    P.G.~Puglia\iEPFL, 
    S.~Dowson\iCCFE,
    M.~Porkolab\iPSFC,
    V.~Guillemot\iESPCI,
    M.~Podest\`{a}\iPPPL,
    M.~Baruzzo\iCCFE,
    R.~Dumont\iCEA, 
    A.~Fasoli\iEPFL,
    M.~Fitzgerald\iCCFE, 
    Ye.O.~Kazakov\iBelgium, 
    M.F.F.~Nave\iLisbon, 
    M.~Nocente\iMilan\iIPST,
    J.~Ongena\iBelgium,
    S.E.~Sharapov\iCCFE,
    \v{Z}.~\v{S}tancar\iSlovenia,
    and JET~Contributors\iJET}
    
    \address{\PSFC \\
             \CCFE \\
             \EPFL \\
             \ESPCI \\
             \PPPL \\
             \CEA \\
             \Belgium \\
             \Lisbon \\
             \Milan \\ \IPST \\
             \Slovenia \\
             \JET}





    \begin{abstract}
    
    The interaction of \AlfvenEigenmodes (\AEs) and energetic particles \yk{is one of many important factors determining} the success of future tokamaks. In JET, eight in-vessel antennas were installed to actively probe \emph{stable} \AEs with frequencies ranging $\SI{25{-}250}{kHz}$ and toroidal mode numbers $\absn < 20$. During the 2019-2020 deuterium campaign, almost 7500 resonances and their frequencies $\fo$, net damping rates $\g < 0$, and toroidal mode numbers were measured in almost 800 plasma discharges. From a statistical analysis of this database, continuum and radiative damping are inferred to increase with edge safety factor, edge magnetic shear, and when including non-ideal effects. Both stable \AE observations and their associated damping rates are found to decrease with $\absn$. Active antenna excitation is also found to be ineffective in \Hmode as opposed to \Lmode; \iaea{this is likely due to the increased edge density gradient's effect on accessibility and ELM-related noise's impact on mode identification}. A novel measurement is reported of a marginally stable, edge-localized Ellipticity-induced \AE probed by the antennas during high-power auxiliary heating (\ICRH and \NBI) up to $\SI{25}{MW}$. \NOVAK kinetic-MHD simulations show good agreement with experimental measurements of $\fo$, $\gamma$, and $\n$, indicating the dominance of continuum and electron Landau damping in this case. Similar experimental and computational studies are planned for the recent hydrogen and ongoing tritium campaigns, in preparation for the upcoming DT campaign.

\end{abstract}
    
\noindent{\it Keywords\/}: \AlfvenEigenmode, stability, Neutral Beam Injection, Ion Cyclotron Resonance Heating 


    \section{Introduction}\label{sec:intro}

    The \AEADiagnostic (\AEAD) actively probes, or excites, \emph{stable} \AlfvenEigenmodes (\AEs) in JET tokamak plasmas%
    \cite{Fasoli1995,Panis2010,Puglia2016}.
    The importance of these \AE stability measurements - i.e. frequencies $\wo = 2\pi\fo$, net damping rates $\g<0$, and toroidal mode numbers $\n$ - cannot be overstated. First, they provide a direct experimental comparison with net growth rates calculated from theory and simulation, from which the contributions of different driving and damping mechanisms can be assessed. Of particular interest is the measurement of alpha particle drive, which is a primary goal of energetic particle (\EP) experiments \cite{Dumont2018} and \AEAD operation in the upcoming JET DT campaign \cite{Joffrin2019}. Importantly, the \AEAD may be the only diagnostic capable of assessing this drive if the alpha population is insufficient to destabilize \AEs. Finally, a better understanding of \AE stability will improve projections of \EP-driven \AEs and the resulting \AE-induced \EP transport in next-step tokamaks, such as ITER and SPARC, and in future fusion pilot plants.

    The JET \AEAD comprises two in-vessel sets of four toroidally spaced antennas positioned below the \nfadd{outboard} midplane and on opposite sides of the torus \cite{Panis2010}. Six amplifiers power six (of eight) antennas with currents ${\sim}\SI{6}{A}$ each; the resulting magnetic perturbation has magnitude $\abs{\delta B/B} < 10^{-3}$ at the plasma edge. Independent phasing of the antennas allows power to be injected into a spectrum of toroidal mode numbers, $\absn < 20$ \cite{Puglia2016}. As the scanning antenna frequency passes through that of a stable \AE, the plasma resonates like a driven, weakly damped harmonic oscillator \cite{Fasoli1995}, and the frequency-filtered magnetic response - obtained from a toroidal array of fast magnetic probes - determines $\fo$, $\g$, and $\n$ \cite{Tinguely2020}. \nfadd{The external antennas are more likely to excite \AEs near the plasma edge than in the core; these often include Global, Toroidicity-induced, and Ellipticity-induced \AEs (\GAEs, \TAEs, and \EAEs, respectively) \cite{Tinguely2021}.}

    This paper reports on recent progress in experimental and computational studies of \AE stability with the \AEAD. The organization is as follows: \cref{sec:motivation} gives a brief review of past results and motivates this study. Then, in \cref{sec:database}, an expanded database of \AE and plasma parameters is presented, along with statistically significant trends related to \AE physics. \Cref{sec:experiment} focuses on the novel measurement of a stable \AE probed by the \AEAD during high-power external heating of a D-$\Hethree$ plasma, and experimental results are compared with kinetic-MHD simulations in \cref{sec:simulation}. Finally, a summary is given in \cref{sec:summary}.

    \section{Motivation}\label{sec:motivation}
    
    Many past studies have analyzed stable \AEs in JET with the \AEAD%
    \cite{Fasoli1995,Fasoli1995nf,Fasoli1996,Fasoli1997,Heidbrink1997,Jaun1998,Wong1999,Fasoli2000,Fasoli2000pla,Jaun2001,Testa2001,Fasoli2002,Testa2003,Testa2003NBI,Testa2003rsi,Testa2004,Testa2005,Testa2006,Fasoli2007,Klein2008,Fasoli2010,Panis2010,Testa2010,Testa2010epl,Testa2011,Testa2011fed,Panis2012a,Panis2012b,Testa2012,Testa2014,Puglia2016,Nabais2018,Aslanyan2019,Tinguely2020,Tinguely2021},
    including measurements made during the 1997 JET DT campaign \cite{Fasoli2000}. However, at the time, the saddle coil system was ineffective in probing stable \AEs during high performance phases in \xpoint magnetic configuration \cite{Fasoli1997,Fasoli2000,Testa2001,Fasoli2002}. 
    Since then, there have been upgrades to the system \cite{Panis2010}, including the independent powering and phasing of the antennas \cite{Puglia2016}, with the main aim of measuring \AE stability in the upcoming JET DT campaign \cite{Joffrin2019}. Thus, it is of interest to map the operational space of the current system, compare with past experimental results, and optimize for near-future studies.
    
    The focus of this paper is the demonstration of the \AEAD's ability to excite stable \AEs in high-power, high-performance plasmas. First, trends in the net damping rate with auxiliary heating and confinement regime are presented in the next section. Then, in \cref{sec:experiment,sec:simulation}, a novel \AE stability measurement is investigated in depth for one high-power D-$\Hethree$ plasma discharge. The observation of this stable \AE was actually surprising in several ways: First, significant ion Landau damping was expected from the Neutral Beam Injection (\NBI) heating power $\Pnbi \approx \SI{20}{MW}$ during the discharge since significant \NBI damping of \AEs has been observed before in many devices, including JT-60U\cite{Saigusa1997}, TFTR \cite{Wong1999}, and JET \cite{Fasoli1997,Testa2003NBI,Dumont2018}. The plasma was also in \xpoint configuration, which makes stable \AE detection even less likely \cite{Dvornova2020,Tinguely2021}. Yet \nfadd{we were able to track a stable \AE in real time, during \xpoint configuration, and at the highest external heating power to-date (${\sim}\SI{25}{MW}$).} Such a measurement gives the authors confidence in overcoming evidence of limited diagnostic efficiency for successful operation in DT.

    \section{Database studies of stable \AlfvenEigenmodes}\label{sec:database}

    This section presents a statistical analysis of thousands of stable \AEs collected in the recent 2019-2020 JET deuterium campaign. The \AEAD was operated on almost 800 plasma discharges, from which a database of almost 7500 stable \AE measurements was assembled. Note that this database is actually an expansion of that reported earlier in \cite{Tinguely2020} and \cite{Tinguely2021} because new data were acquired after their publications. 

    In the following analyses and unless otherwise stated, data are restricted to normalized damping rates $-\go \leq 6\%$, uncertainties $\Delta \fo < \SI{1}{kHz}$ and $\dgo < 1\%$, \xpoint magnetic configuration, and heating powers $\Pnbi < \SI{7}{MW}$ and $\Picrh < \SI{7}{MW}$ from Ion Cyclotron Resonance Heating (\ICRH). The last two bounds are motivated by the sometimes-overwhelming pick-up in magnetics data which can lead to misidentification of \AEs \cite{Fasoli2010}.  Thus, resonances observed during $\Pnbi, \Picrh > \SI{7}{MW}$ require closer inspection, as done in \cref{sec:experiment,sec:simulation}.
    
    \subsection{Statistical analyses of damping rates}
    
        Two recent works have analyzed stable \AEs from this database: The damping rate was observed to increase linearly with the edge safety factor $\qnf$ \nfadd{(see Fig.~5a in \cite{Tinguely2020})}. This was attributed to an increase in continuum damping as the \AE continuum gap closes at the plasma edge, which was also seen before in JET \cite{Panis2012b}. The damping rate was also noted to increase rapidly and nonlinearly with the edge magnetic shear $\snf$ \nfadd{(see Fig.~6a in \cite{Tinguely2021})}, likely due to both continuum and radiative damping. This result was in agreement with past works \cite{Fasoli1997,Fasoli2000,Testa2001,Fasoli2002,Testa2005,Panis2012b}. These trends are confirmed to have statistical significance in \cref{tab:correlations}, which reports the linear correlation $\rw(-\go,x)$ of the normalized damping rate, weighted by its inverse variance $\dgo^{-2}$, with various parameters $x$. Here, magnitudes $\abs{\rw} \geq 0.5$ are considered significant.
        
        \begin{table}[h!]
            \centering
            \caption{Weighted linear correlations of normalized damping rate. Subscripts $0$ and $95$ refer to parameters evaluated at normalized poloidal flux values $\psin = 0$ and $0.95$, respectively. Note that the correlation with \NBI heating power is restricted to values $\Pnbi > 0$ during no \ICRH ($\Picrh = 0$), and vice versa. Bold values, $\abs{\rw} \geq 0.5$, are considered significant.}
            \label{tab:correlations}
            \begin{tabular}{c c}
                \hline
                Parameter $x$   & $\rw(-\go,x)$ \\
                \hline
                $\qo$       & $-0.11$ \\
                $\qnf$      & $\mathbf{0.54}$ \\
                $\snf$      & $\mathbf{0.57}$ \\
                $\lambda$   & $\mathbf{0.69}$ \\
                $\Bo$       & $-0.12$ \\
                $\Teo$      & $-0.01$ \\
                $\neo$      & $-0.20$ \\
                $\nnf$      & $-0.34$ \\
                $\gradnnf$  & $0.29$ \\
                $\Pnbi$     & $0.28$ \\
                $\Picrh$    & $-0.09$ \\
                \hline
            \end{tabular}
        \end{table}
        
\begin{figure}[h!]
            \centering
            \begin{subfigure}{\halfwidth}
                \includegraphics[width=\textwidth]{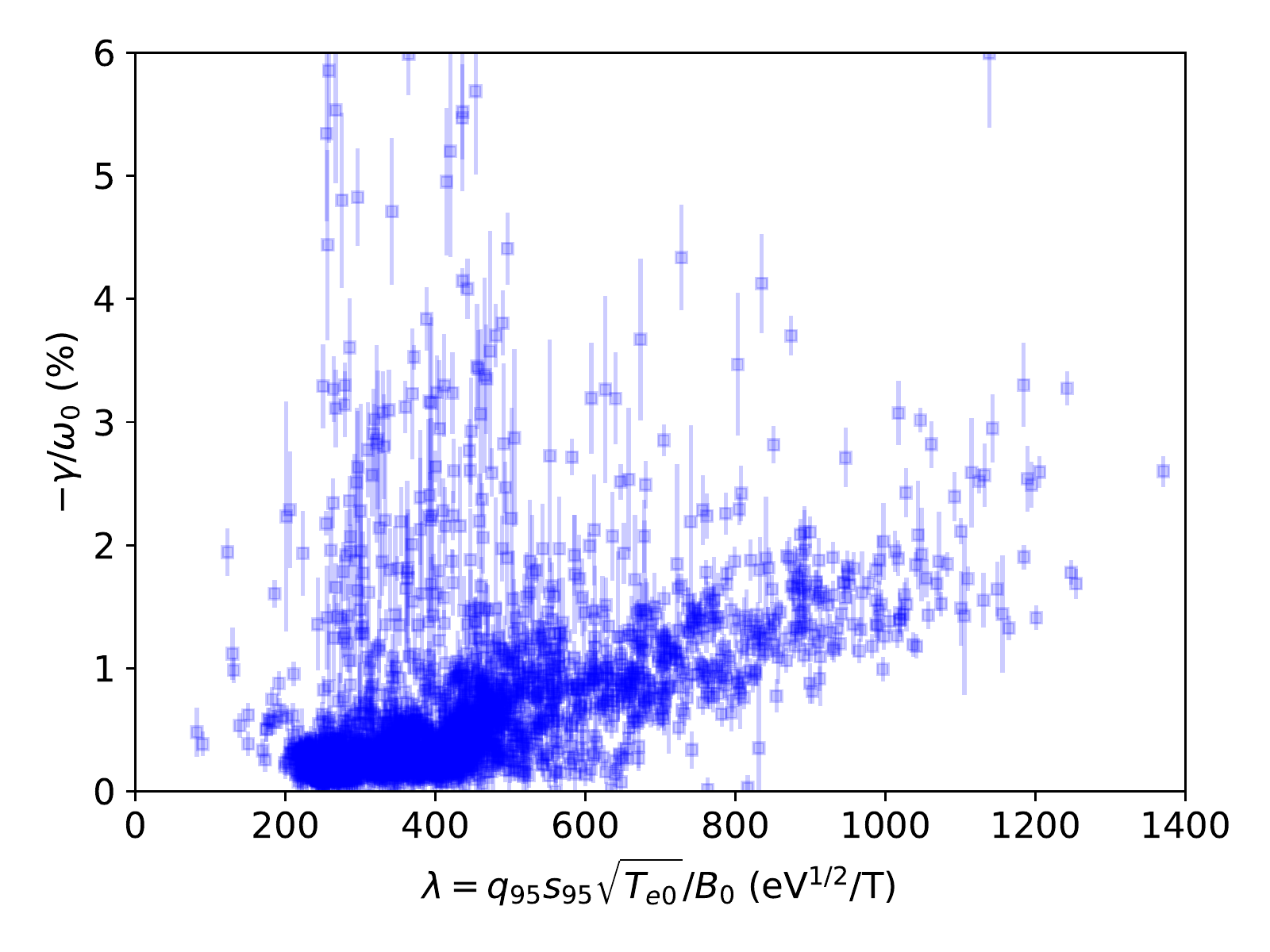}
                \caption{}
                \label{fig:lambda}
            \end{subfigure}
            \begin{subfigure}{\halfwidth}
                \includegraphics[width=\textwidth]{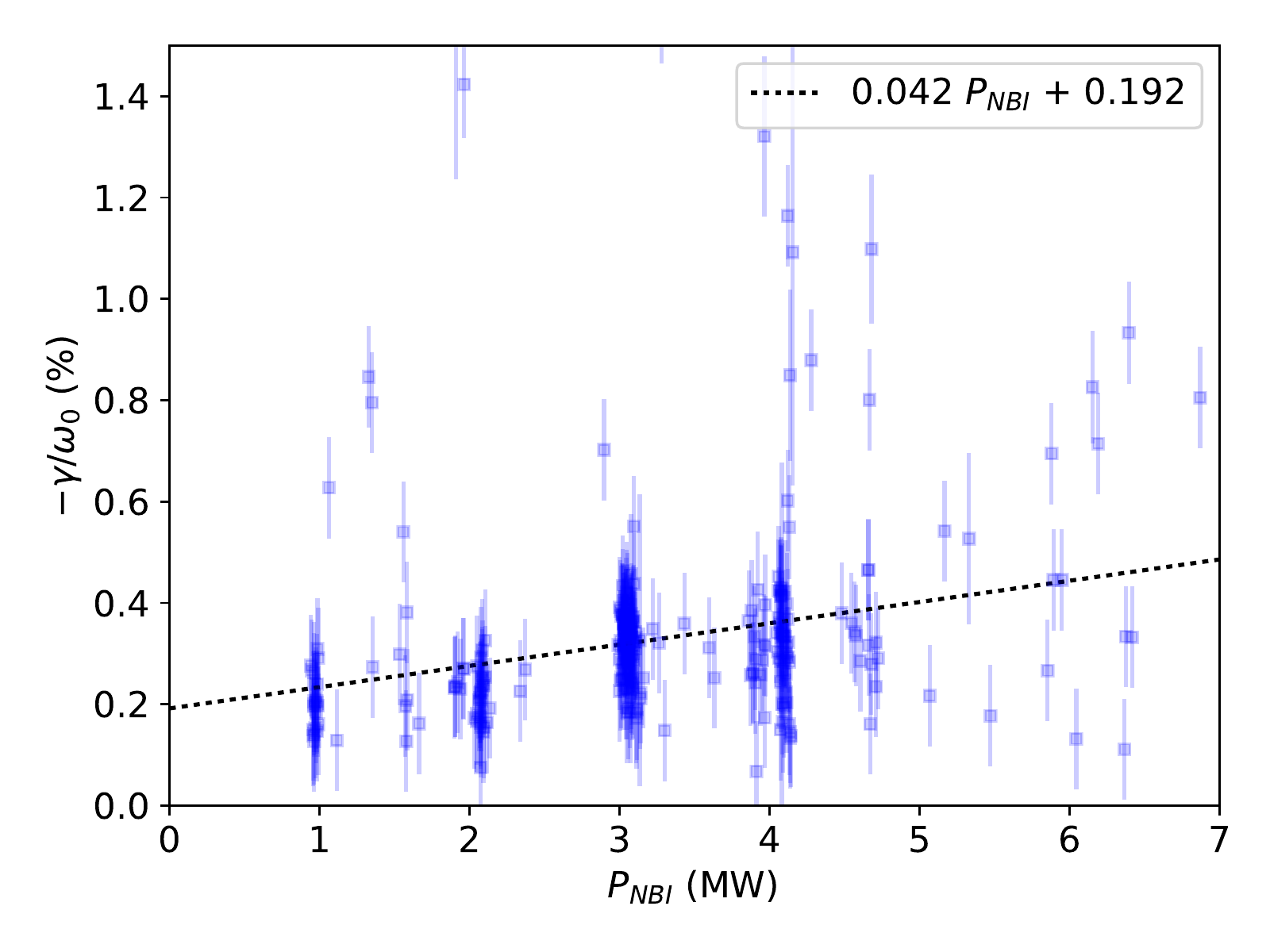}
                \caption{}
                \label{fig:nbi}
            \end{subfigure}
            \caption{Normalized damping rate vs (a)~non-ideal parameter and (b)~\NBI power, with uncertainties as error bars. A linear fit is overlaid in (b).}
        \end{figure}

        The highest correlation is with the so-called non-ideal parameter $\lambda = \qnf\snf\sqrt{\Teo}/\Bo$ \cite{Heidbrink2008,Panis2012a}, where $\Teo$ and $\Bo$ are the on-axis values of the electron temperature and toroidal magnetic field, respectively. Note that no correlation is observed with  $\Teo$ and $\Bo$ individually. The non-ideal parameter is key in the theory of radiative damping \cite{Mett1992,Mett1994}, and therefore $\lambda$ is a better indicator of its impact than $\qnf$ or $\snf$ alone.
        \iaea{Previous computational efforts with a variety of MHD, kinetic, and gyrokinetic codes \cite{Borba2010} had also identified enhanced radiative damping with increasing temperature and hence larger gyro-radius effects.}
        All data points are shown in \cref{fig:lambda} \nfadd{with partial transparency}, \nfsub{where} \nfadd{and} a clear linear trend is observed \nfadd{in the highest-density (i.e. the darkest) region}. \nfadd{At present, the divergence from this trend at low $\lambda$ ($< \SI{600}{eV^{1/2}/T}$) is not well understood.}
        
        
        Correlations with \NBI and \ICRH powers are also given in \cref{tab:correlations}. Here, data are restricted to non-zero values \nfadd{of heating power} since wide variation in the damping rate is observed with no external heating. Additionally, only one external heating source is considered at a time, i.e. $\Pnbi > 0$ when $\Picrh = 0$, and vice versa. Even with these restrictions, there are still hundreds of data points to analyze. \nfsub{While the reported correlations are relatively poor, linear fits} \nfadd{While the reported correlation with $\Picrh$ is poor, that with $\Pnbi$ is moderate, and a linear fit} to the damping rate indicates increased damping with \NBI (also observed in \cite{Testa2003NBI})\nfsub{ and decreased damping with \ICRH (also observed in \cite{Fasoli1997})}. This is expected on JET because \NBI fast ion energies are ${<}\SI{100}{keV}$, which is typically lower than the \Alfven speed $\vA$.\nfsub{, whereas \ICRH-accelerated fast ions often have higher energies, sometimes ${>}\SI{1}{MeV}$.}
        
        
        Damping rate data vs \NBI power are shown in \cref{fig:nbi}, with clusters around $\Pnbi \approx \SI{1,2,3,4}{MW}$ resulting from the discrete steps in power delivered by individual injectors. The calculated slope is low, only 0.042\%/MW, extrapolating to $-\go \approx 1.5\%$ at $\Pnbi \approx \SI{30}{MW}$. Because the \AEAD primarily probes stable \AEs localized near the plasma edge while \NBI power is mostly deposited in the core, this extrapolation is realistically a lower bound on \NBI ion Landau damping. \nfadd{Yet, it is important to note that there is sufficient scatter in the data, and the trend could be caused by conflating factors, though not $\Picrh$.}
        
    \subsection{Trends related to operational scenarios}
    
        Investigations of the \AEAD operational space are also important, especially as they relate to high-power, high-performance scenarios and the upcoming JET DT campaign. In \cite{Tinguely2020}, it was reported that the \AEAD's efficiency of resonantly exciting stable \AEs decreased with increasing plasma current and external heating power. A thorough study of \AEAD-plasma coupling \cite{Tinguely2021} also showed reduced efficiency in \xpoint (i.e. diverted) magnetic configuration compared to limiter configuration - consistent with the $\qnf$ and $\snf$ trends above - as well as with increased plasma-antenna separation.
        
        Recently, the EUROfusion JET-ILW pedestal database \cite{Frassinetti2020} was compared with the stable \AE database: Interestingly, of the approximately 500 stable \AEs measured in over 80 plasma discharges common to both databases, \emph{none} were observed during \Hmode periods. This confirms, with relatively high confidence (p-value = 0.076), that the \AEAD \iaea{is inefficient in probing} stable \AEs during \Hmode. One probable cause is the effect of the density pedestal on the \AE continuum. In fact, recent modeling work \cite{Dvornova2020} highlighted the impact of the edge density profile on \AE continua as well as \AEAD accessibility. Yet, there still exist some stable \AE measurements during high performance operation, as explored in \cref{sec:experiment}.
        
        
        To investigate further the effect of edge density conditions - and complement that of edge magnetic conditions $\qnf$ and $\snf$ above - the edge electron density $\nnf = \ne(\psin=0.95)$, obtained from Thomson Scattering, and its gradient $\gradnnf = \rd \ne / \rd \psin \vert_{\psin=0.95}$ were also analyzed, with $\psin$ the normalized poloidal flux. While no ``good'' correlations, i.e. $\abs{\rw} \geq 0.5$, are identified (see \cref{tab:correlations}), the strongest is found with $\nnf$. 
        
\begin{figure}[h!]
            \centering
            \begin{subfigure}{\halfwidth}
                \includegraphics[width=\textwidth]{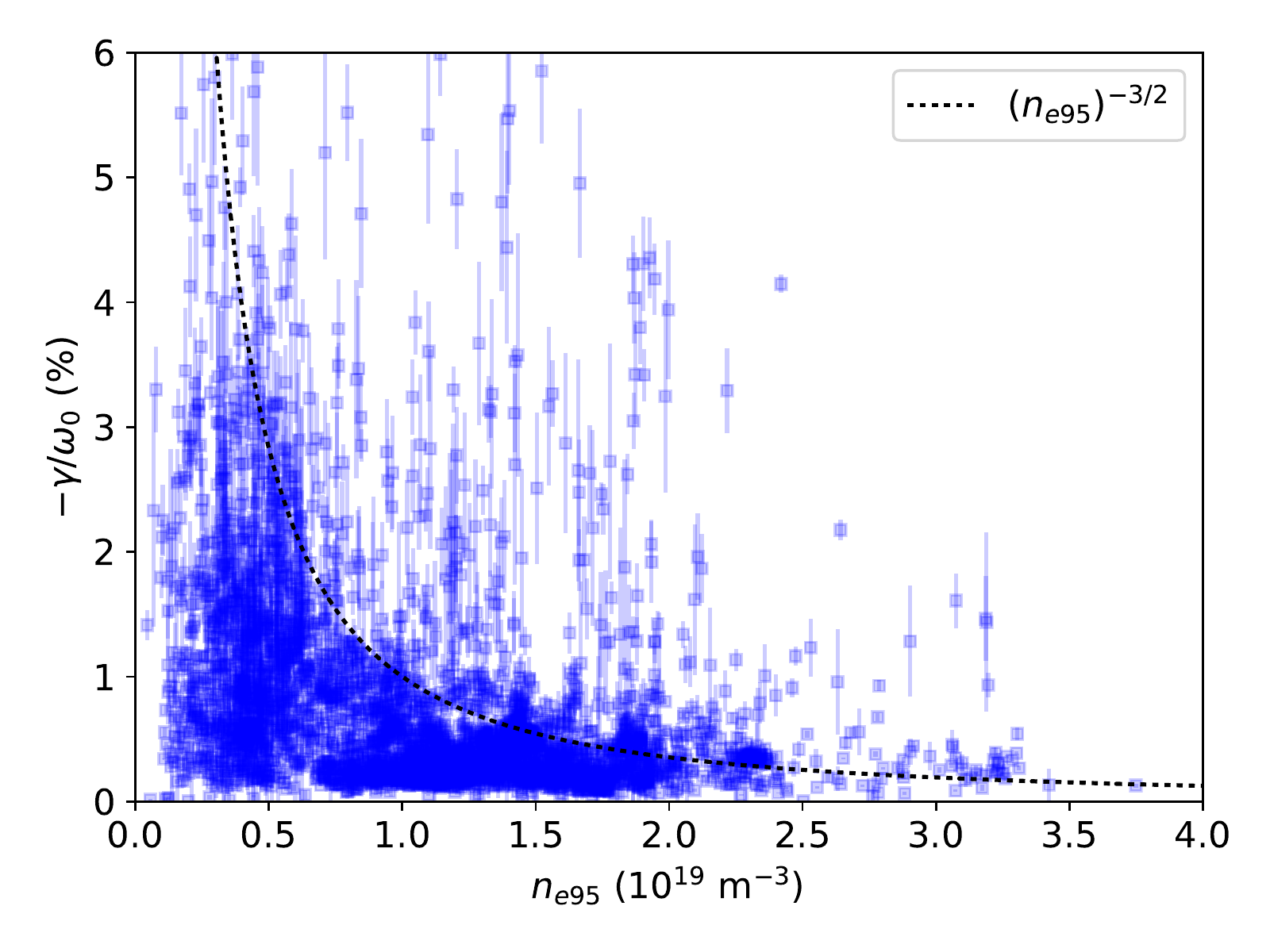}
                \caption{}
                \label{fig:n95}
            \end{subfigure}
            \begin{subfigure}{\halfwidth}
                \includegraphics[width=\textwidth]{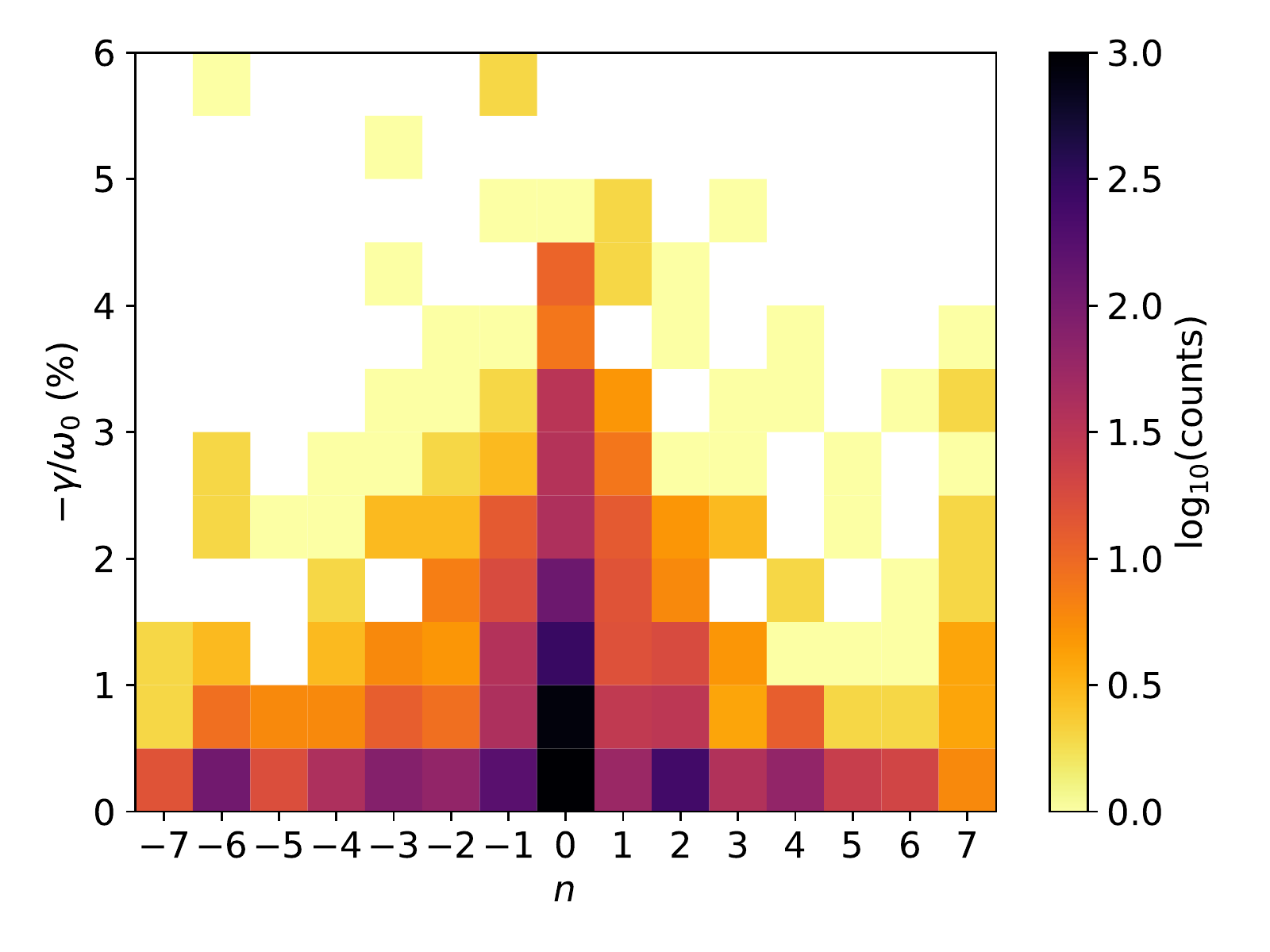}
                \caption{}
                \label{fig:n}
            \end{subfigure}
            \caption{(a)~Normalized damping rate vs edge electron density $\nnf$ with an approximate bounding curve overlaid. (b)~Number of stable \AE observations (logarithmic) vs toroidal mode number ($\absn \leq 7$) and normalized damping rate, with uncertainties restricted to $\dgo \leq  0.5\%$.}
        \end{figure}

        All data points are shown versus $\nnf$ in \cref{fig:n95}. Interestingly, there is a clear upper bound on the data, $ -\go \leq  (\nnf)^{-3/2}$, with ${\sim}84\%$ of data falling below this curve.%
            \footnote{\nfadd{Note that the scaling of this curve is arbitrary.}}
        This trend is counterintuitive as the \AE gap is expected to close with \emph{increasing} $\nnf$ via $\vA \propto \ne^{-1/2}$, here assuming $\ne \approx \ni$, so that enhanced continuum damping might be observed.%
        \footnote{\iaea{This also depends on the safety factor profile.}}
        However, note that gradient of the continuum contains the term $\rd \vA / \rd \ne  \propto \ne^{-3/2}$, matching the bounding curve. \nfsub{Though not shown here, a linear upper bound is found with $\gradnnf$.} This suggests that as $\nnf$ increases \nfadd{(}and $\gradnnf$ becomes steeper\nfadd{)}, \AEAD accessibility is reduced and stable \AEs become more difficult to excite, consistent with no observations in \Hmode as discussed above.
        
        
        \iaea{It is important to note that reduced \AEAD accessibility is not the only issue here; the \emph{detection} of stable \AEs by the fast magnetic probes is also reduced in \Hmode. During high-power external heating, significant noise is seen in the magnetics signals \nfadd{(e.g. from ELMs)}, making stable \AEs difficult to identify, even if the \AEAD is able to resonate with them. \nfsub{Though not shown here, it is observed that the probability of resonance detection decreases as the ELM frequency increases.} Thus, there are several conflating factors conspiring against successful \AEAD operation during \Hmode periods.}
        
        
        Finally, the dependencies of the damping rate on toroidal mode number are explored. \Cref{fig:n} shows the number of stable \AE measurements for each estimated $\n$ and bin of measured $\go$. Here, data are restricted to $\absn \leq 7$, consistent with typical \emph{unstable} \AEs observed in JET, and $\abs{\dgo} < 0.5\%$, allowing a finer grid discretization. As discussed in \cite{Tinguely2020}, the high density of $\n = 0$ values could indicate that these are true Global \AEs (\GAEs) or that too few magnetic probes were available for a good $\n$ estimation.
        
        A general trend of decreasing $\abs{\go}$ with $\absn$, also noted in \cite{Tinguely2020}, is observed in \cref{fig:n}. There are several possible explanations: For a given \AE radial location, the mode width decreases as $1/\absn$ leading to more localized damping, as opposed to more global modes interacting with, say, the continuum. Also, in the presence of fast ions (\FI), \AE drive increases with $n$ via $\n\wfi \propto (\n/r)  \rd \pfi/\rd r$, the \FI radial pressure gradient. However, an asymmetry for positive and negative $\n$ would be expected, which is not easily observed in \cref{fig:n}. Unfortunately, even within a database of almost 7500 stable \AEs, no $\pm\n$ pair exists with sufficiently similar plasma or \FI conditions to estimate $\rd \pfi/\rd r$, as suggested in \cite{Fasoli2002}.

    \section{Experimental measurements of marginal \AE stability during high-power heating}\label{sec:experiment}

    This section reports on a novel measurement of \AE stability during JPN~94703 with high auxiliary heating power, $\Pnbi+\Picrh \approx \SI{25}{MW}$. This pulse was part of the three-ion-heating scenario development experiments at JET relevant to the upcoming JET DT campaign as well as \ICRH in ITER \cite{Joffrin2019,Kazakov2020,Nocente2020}. Experimental results are presented here, and comparisons with kinetic-MHD simulations are given in \cref{sec:simulation}.

\begin{figure}[h!]
        \centering
        \begin{subfigure}{\halfwidth}
            \includegraphics[width=\textwidth]{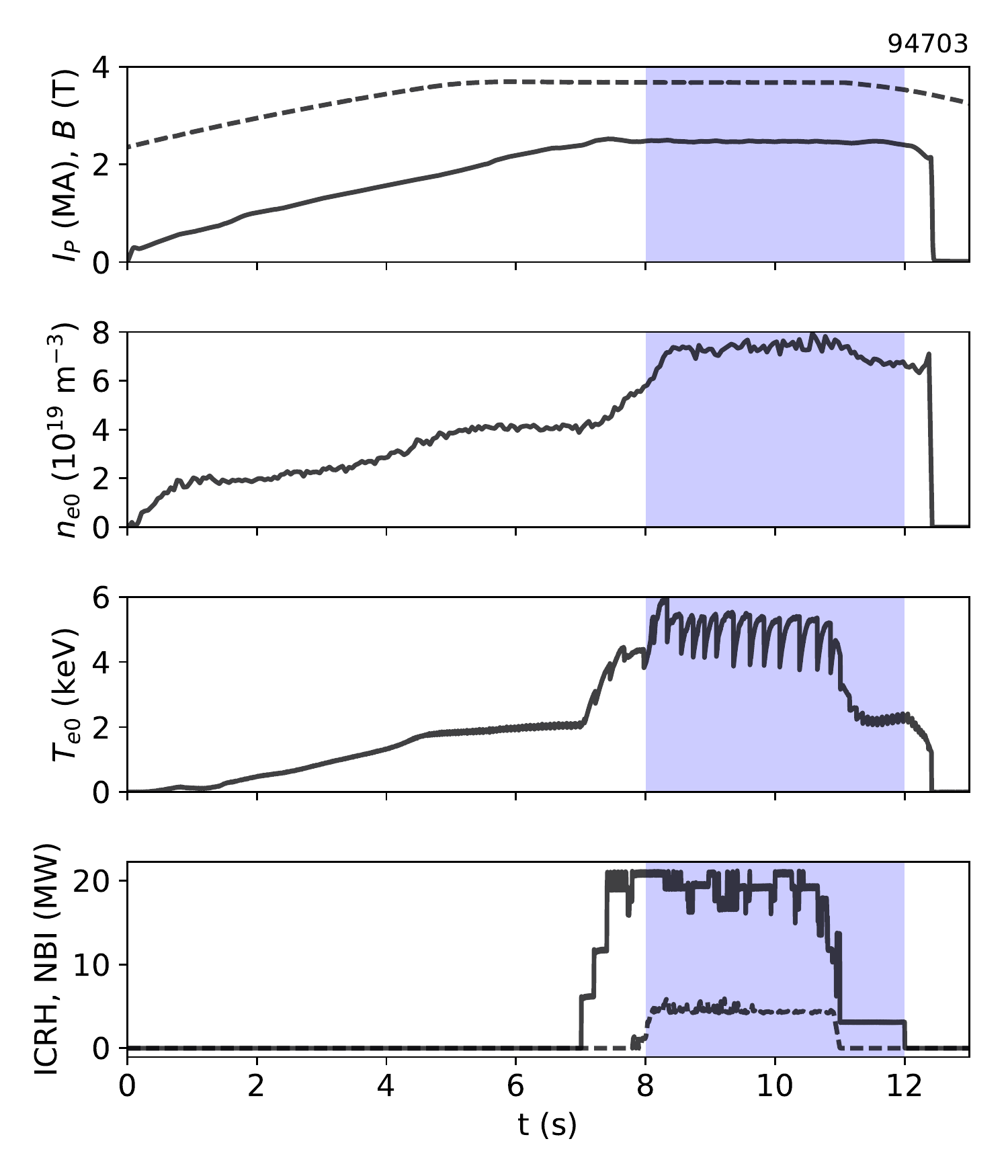}
            \caption{}
            \label{fig:params}
        \end{subfigure}
        \begin{subfigure}{\halfwidth}
            \includegraphics[width=\textwidth]{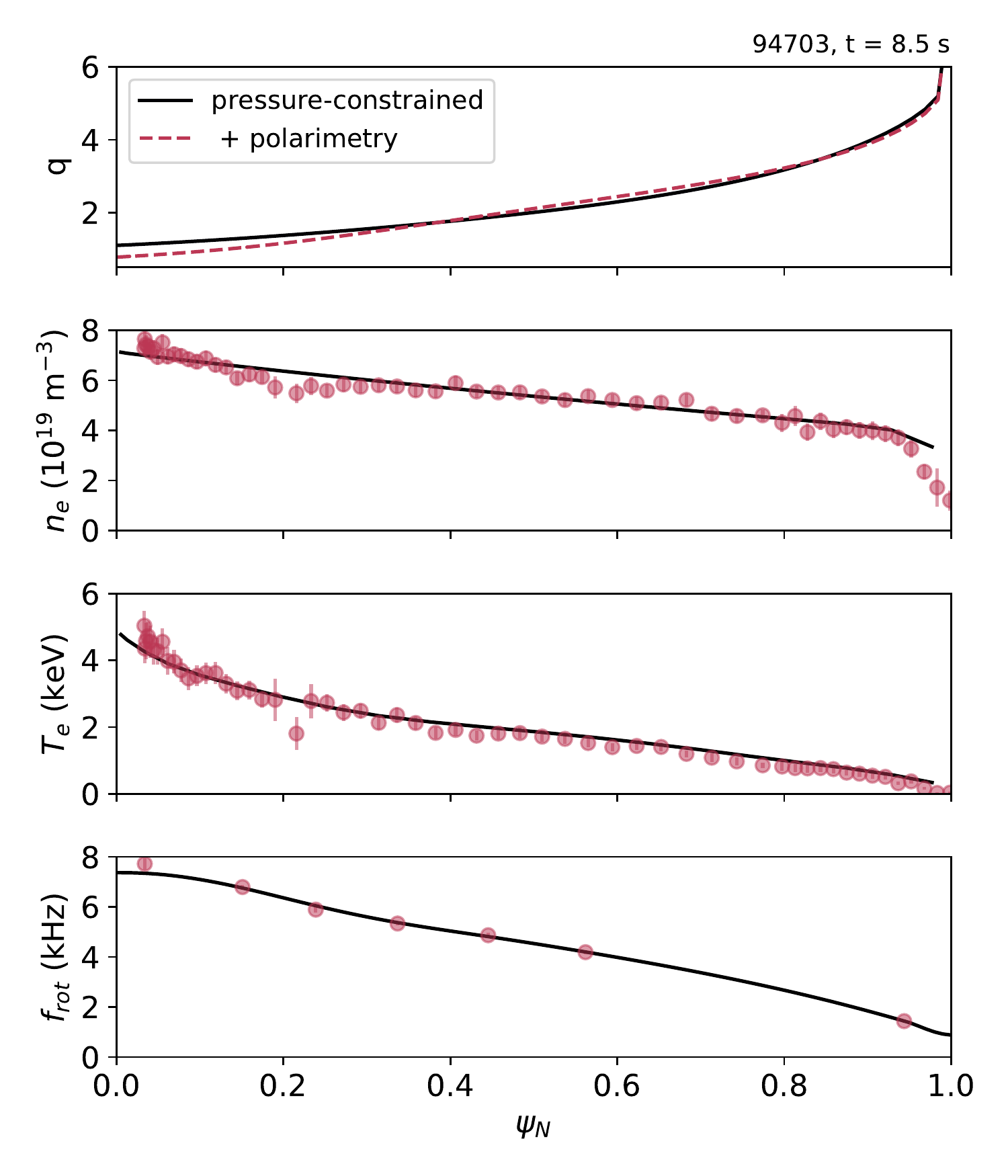}
            \caption{}
            \label{fig:profiles}
        \end{subfigure}
        \caption{(a)~Plasma parameters for JPN~94703: toroidal magnetic field (dashed), plasma current (solid), central electron density from Thomson Scattering (TS) and temperature from Electron Cyclotron Emission, and heating powers from \NBI (solid) and \ICRH (dashed). A stable \AE was tracked during the shaded time interval. (b)~Profiles at $t = \SI{8.5}{s}$: safety factor from \EFIT constrained by pressure  only (solid) and additionally polarimetry (dashed), electron density and temperature from TS, and rotation frequency from $\Hethree$ charge exchange. Experimental data are shown as circles with uncertainties as error bars, while solid lines are fitted data.}
        \label{fig:params_and_profiles}
    \end{figure}

    Time traces of plasma parameters for JPN~94703 are shown in \cref{fig:params}, with the time range of interest, $t = \SI{8{-}12}{s}$, shaded. Flattop parameters are $\Bo = \SI{3.7}{T}$, $\Ip = \SI{2.5}{MA}$, $\neo \approx \SI{8\times10^{19}}{m^{-3}}$, and $\Teo \approx \SI{5.3}{keV}$. Auxiliary heating are $\Pnbi \approx \SI{19{-}21}{MW}$ and $\Picrh \approx \SI{4.4}{MW}$ from $\t = \SI{8{-}11}{s}$. The concentration of $\Hethree$ is relatively high, $\nHethree/\ne \approx 23\%$, as part of the D-$\DNBI$-$\Hethree$ heating scheme \cite{Kazakov2021}.
    
    Plasma profiles are shown for one time, $\t = \SI{8.5}{s}$, in \cref{fig:profiles} as a function of the normalized poloidal flux $\psin$. Two $\q$-profiles from \EFIT \cite{Lao1985} are shown: One is constrained by the fitted kinetic profiles from \TRANSP \cite{TRANSP,Hawryluk1980,Ongena2012}, which match Thomson Scattering $\ne$ and $\Te$  data well. (Here, equal electron and ion temperatures, $\Te = \Ti$, are assumed.) The other $q$-profile is additionally constrained by polarimetry measurements and agrees within ${\sim}10\%$. \yk{Note that this time-slice is between two $\Te$ sawtooth crashes at $\t \approx \SI{8.33}{s}$ and $\SI{8.55}{s}$, and electron cyclotron emission data indicate that the inversion radius is at $R\approx\SI{3.13}{m}$ ($\psin \approx 0.03$). Thus, the $\q$-profile fit with polarimetry is likely \nfsub{``more correct''} \nfadd{more accurate}; however, as will be seen in the next section, the \AE analysis is fairly robust to these uncertainties in $q$.} Finally, rotation data is obtained from $\Hethree$ charge exchange spectroscopy. 
    

\begin{figure}[h!]
        \centering
        \begin{subfigure}{\halfwidth}
            \includegraphics[width=\textwidth]{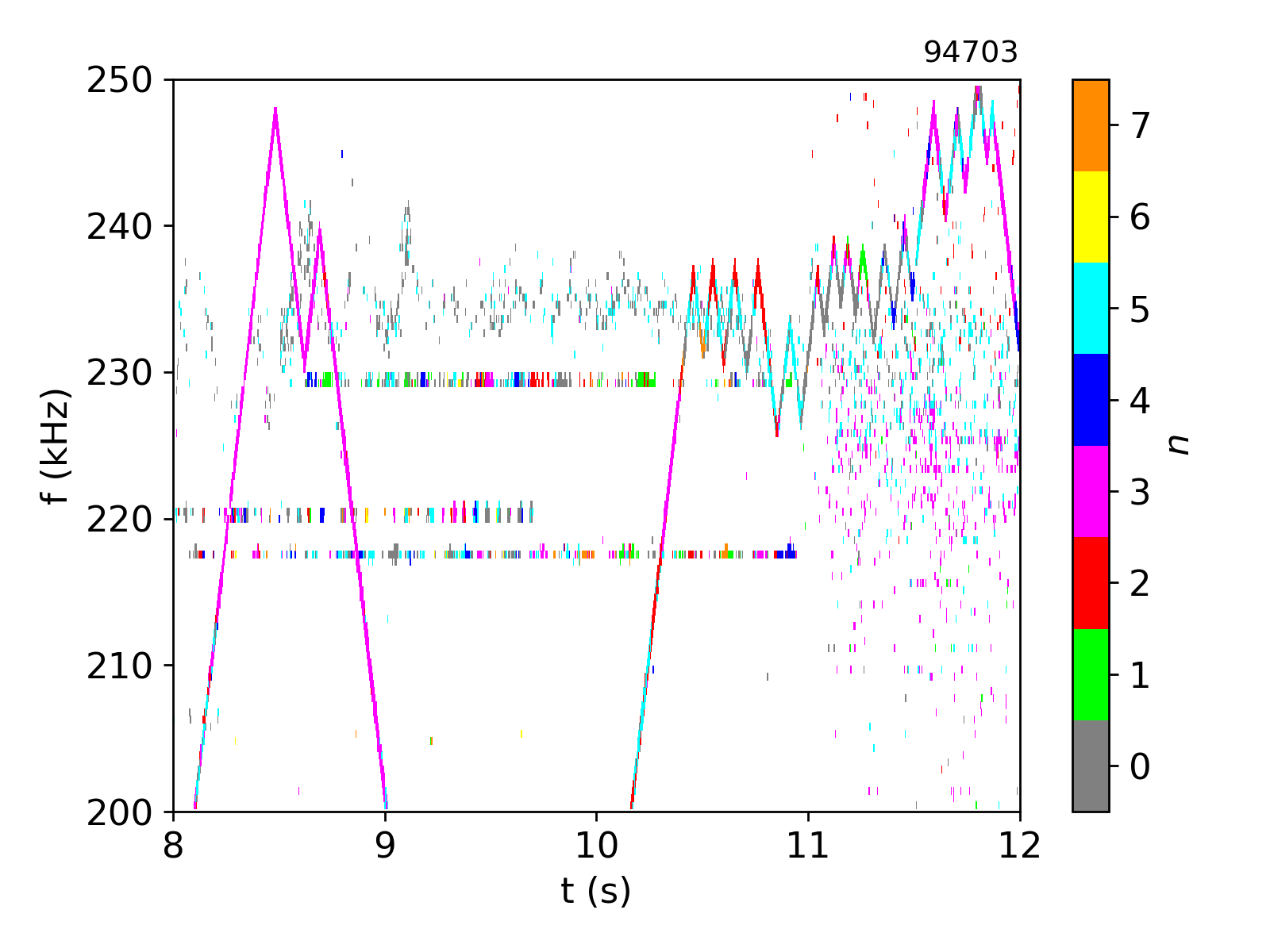}
            \caption{}
            \label{fig:spectrogram}
        \end{subfigure}
        \begin{subfigure}{\halfwidth}
            \includegraphics[width=\textwidth]{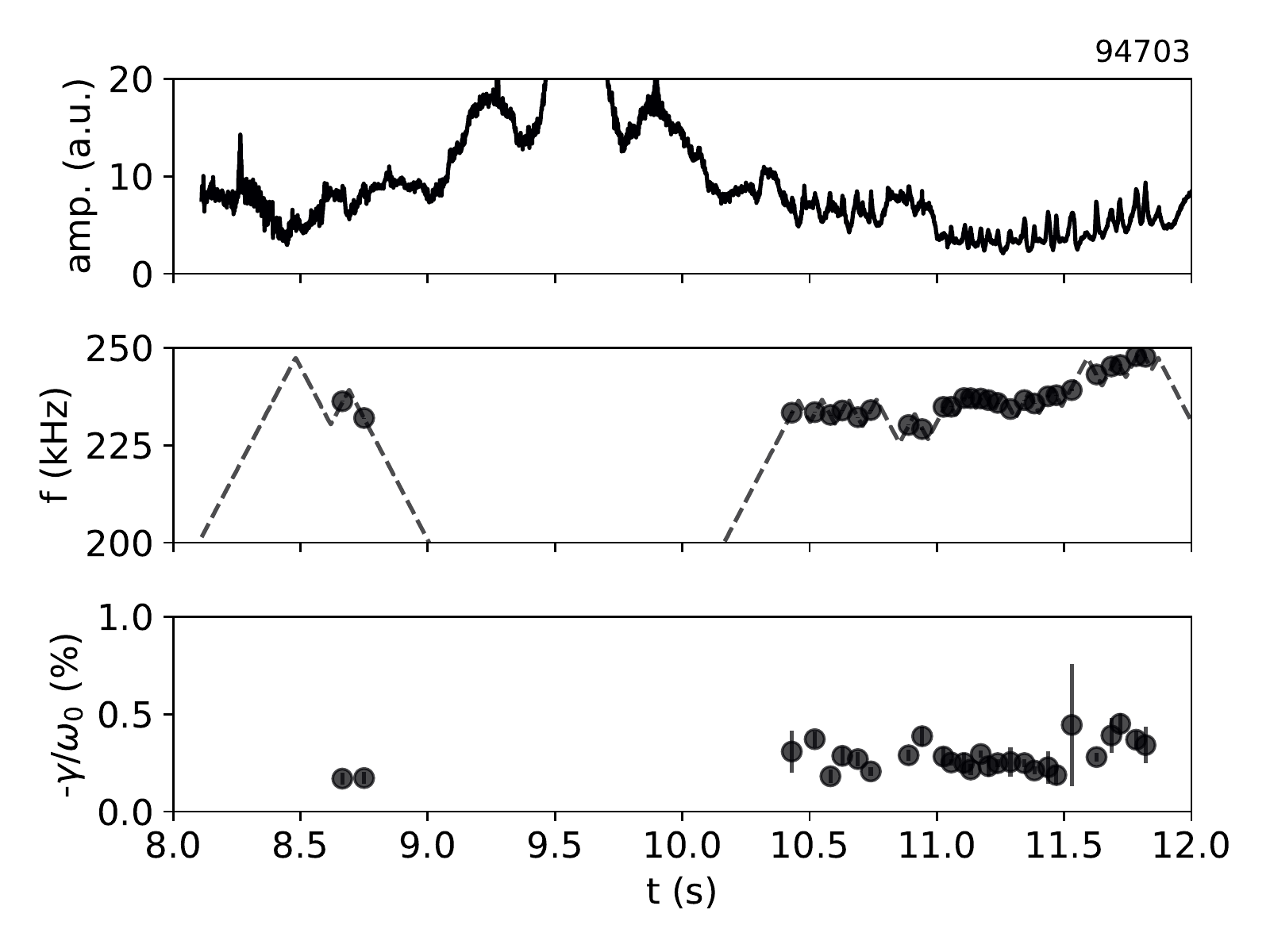}
            \caption{}
            \label{fig:resonances}
        \end{subfigure}
        \caption{(a)~Fourier decomposition of magnetics data with toroidal mode number analysis. (b)~Stable \AE resonance measurements: the magnetic response amplitude (summed over all probes), \AEAD (dashed) and resonant frequencies (circles), and normalized damping rates with uncertainties as error bars.}
        \label{fig:spectrogram_and_resonances}
    \end{figure}

    A stable \AE was tracked in real time during the high heating power phase. This can be seen in the Fourier \nfsub{analysis} \nfadd{decomposition} of magnetic signals in \cref{fig:spectrogram}. The triangular waveform is the scanning \AEAD frequency, $\f = \SI{125{-}250}{kHz}$. The antenna phasing was such that power was injected primarily into odd toroidal mode numbers (i.e. $\absn = 1, 3, 5,\dots$) which is confirmed by the mode number analysis of toroidally distributed magnetic probes' data, showing primarily $\n = 3$ (magenta). Around $\t = \SI{8.5}{s}$, a stable \AE is detected \nfadd{by the \AEAD} at $\fo \approx \SI{235}{kHz}$, \nfadd{and the real-time monitoring system quickly changes the scan direction to track the mode.} \nfsub{but is then quickly lost by the real-time tracking system.} \nfadd{Unfortunately, tracking is ``lost'' just before $\t \approx \SI{9}{s}$, and the scan continues downward in frequency.} A marginally \emph{unstable} \AE is then seen between $\t \approx \SI{9{-}10}{s}$ with \nfadd{$\fo \approx \SI{235}{kHz}$ and} a mix of $\n = 0$ (grey) and $\n = 5$ (cyan). \nfadd{The fact that the mode is unstable, even with a very small amplitude, could be the reason why the \AEAD did not identify it.} The \AEAD finds the stable mode again at $\t \approx \SI{10.5}{s}$ and then tracks until $\t \approx \SI{12}{s}$ when $\fo \approx \SI{250}{kHz}$.
    

    The measured stable \AE parameters are shown in \cref{fig:resonances}. Here, the same \AE resonance must be detected by at least three (of fourteen) magnetic probes to be considered ``good.'' Characteristic peaks in the magnetic response (summed from all probe amplitudes) are observed, though they are easier to identify by eye after $\t > \SI{10}{s}$ as external heating is lowered. The resonant frequency of the \AE is relatively smooth in time, and \nfadd{the gap in measurements in \cref{fig:resonances} is easily connected by the marginally stable \AE visible in \cref{fig:spectrogram} during that interval.} \nfsub{so is the damping rate which ranges} \nfadd{The damping rate is also relatively constant in time, ranging} from $-\go \approx 0.17\%{-}0.45\%$ with mean value $-\go \approx 0.28\%$ ($-\g \approx \SI{1}{kHz}$), indicating marginal stability. \nfadd{Thus, only a small increase in \AE drive was needed to cause the transition from stability to instability in the interval $\t \approx \SI{9{-}10}{s}$.} \nfadd{Finally,} note how $\go$ does not change much as \NBI power is reduced from $\t \approx \SI{10.5{-}11}{s}$.
    
    \jg{
    Also note in \cref{fig:profiles} that the edge density is $\nnf \approx \SI{4\times10^{19}}{m^{-3}}$, which is higher than all data in the database of the previous section (see 
    \cref{fig:n95}).%
    \footnote{The data from JPN~94703 are not actually in the database of \cref{sec:database} because of the filter $
    \Pnbi < \SI{7}{MW}$.}
    Comparing with the database trend, it is consistent - and perhaps fortuitous - that a marginally stable \AE is observed in this pulse. A stable \AE with higher damping rate may not have been measured at all. In fact, from $\t = \SI{7.5{-}8.6}{s}$, the plasma is in \Mmode \cite{Solano2017}, a weak ELM-free \Hmode state with no temperature pedestal (see \cref{fig:profiles}); the rest of the pulse is confirmed to be in \Lmode. Again, this agrees with the conclusion of the previous section that the \AEAD has difficulty exciting stable \AEs in \Hmode.
    }

    \section{Simulations with the kinetic-MHD code NOVA-K}\label{sec:simulation}

    While the mode number identification of $\n = 5$ seems clear in \cref{fig:spectrogram}, this result is unfortunately quite sensitive to the magnetic probes used. In fact, a separate analysis of the \AE resonances (not shown) returns a range $\absn = 0{-}5$. Additionally, the mode location is difficult to identify because it is stable and thus not seen in the Fourier analysis of interferometry, reflectometry, or soft X-ray data. That said, the \AEAD is known to excite more edge-localized \AEs as an external antenna system \cite{Tinguely2021}. Thus, a range of mode numbers, $\n = 3{-}6$, is simulated with the \NOVAK kinetic-MHD code \cite{Cheng1992,Fu1992,Gorelenkov1999} to assess the existence, mode structure, and stability of \AEs. Input profiles are those at $\t = \SI{8.5}{s}$ (see \cref{fig:profiles}), approximately the time of the first stable \AE measurement (see \cref{fig:resonances}). 

    In addition to various damping mechanisms, \NOVAK also calculates the contribution to the growth rate from \NBI fast ions (\FIs), assuming a slowing down distribution.\footnotemark%
        \footnotetext{The contributions from \ICRH-accelerated fast ions and $\Hethree$ ions are not included in the present work.}
    In JPN~94703, deuterium \NBI ions, with energies ${\sim}\SI{100}{keV}$, were injected via normal and tangential beams with initial pitches around $\vpar/\vo \approx 0.44$ and $0.62$, respectively. The deuterium \FI distribution function is computed in \TRANSP using the NUBEAM and TORIC modules, with \NBI and \ICRH synergy accounted through the Monte Carlo kick model. 
    
\begin{figure}[h!]
        \centering
        \begin{subfigure}{\halfwidth}
            \includegraphics[width=\textwidth]{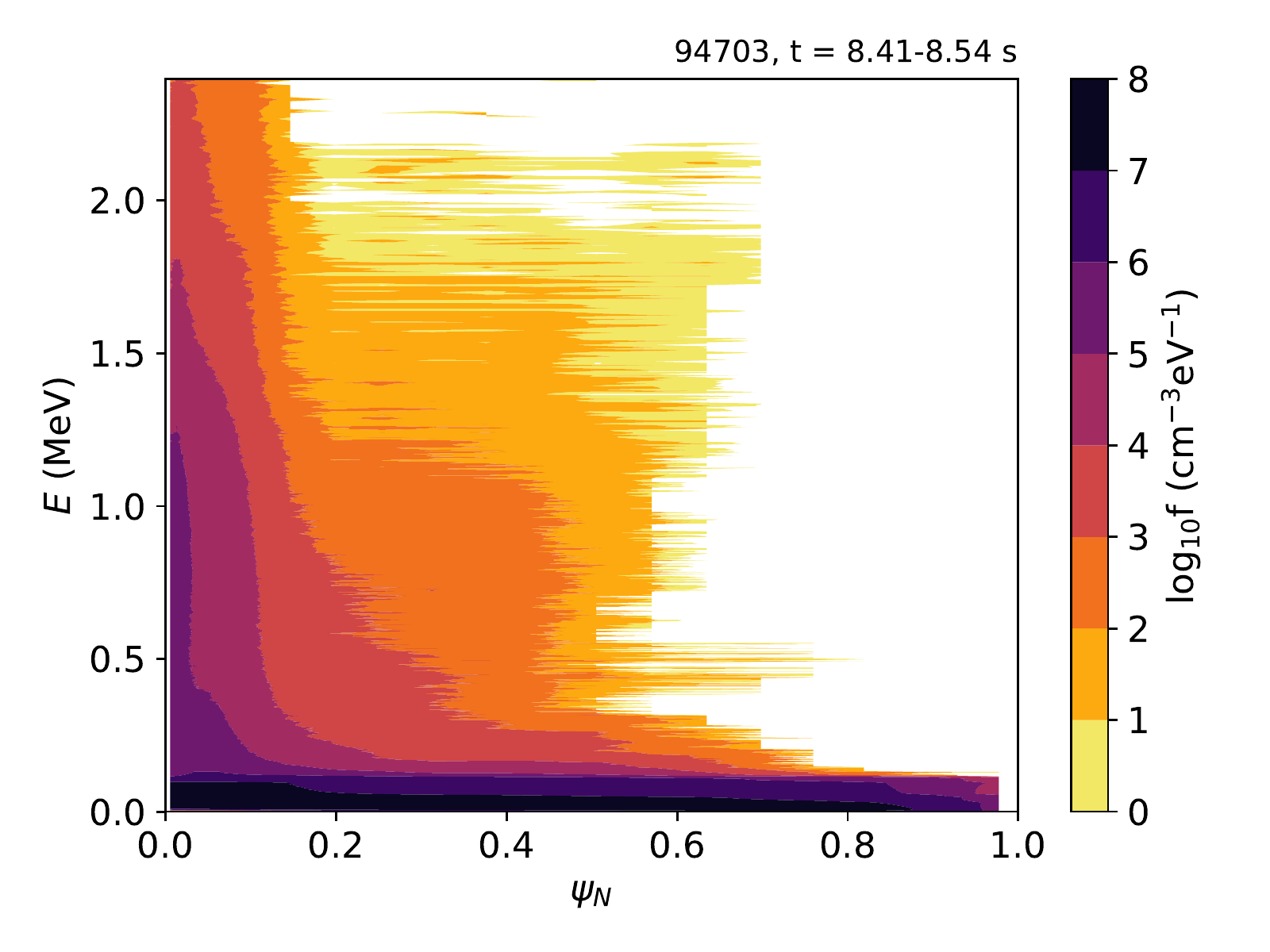}
            \caption{}
            \label{fig:fidf_psiE}
        \end{subfigure}
        \begin{subfigure}{\halfwidth}
            \includegraphics[width=\textwidth]{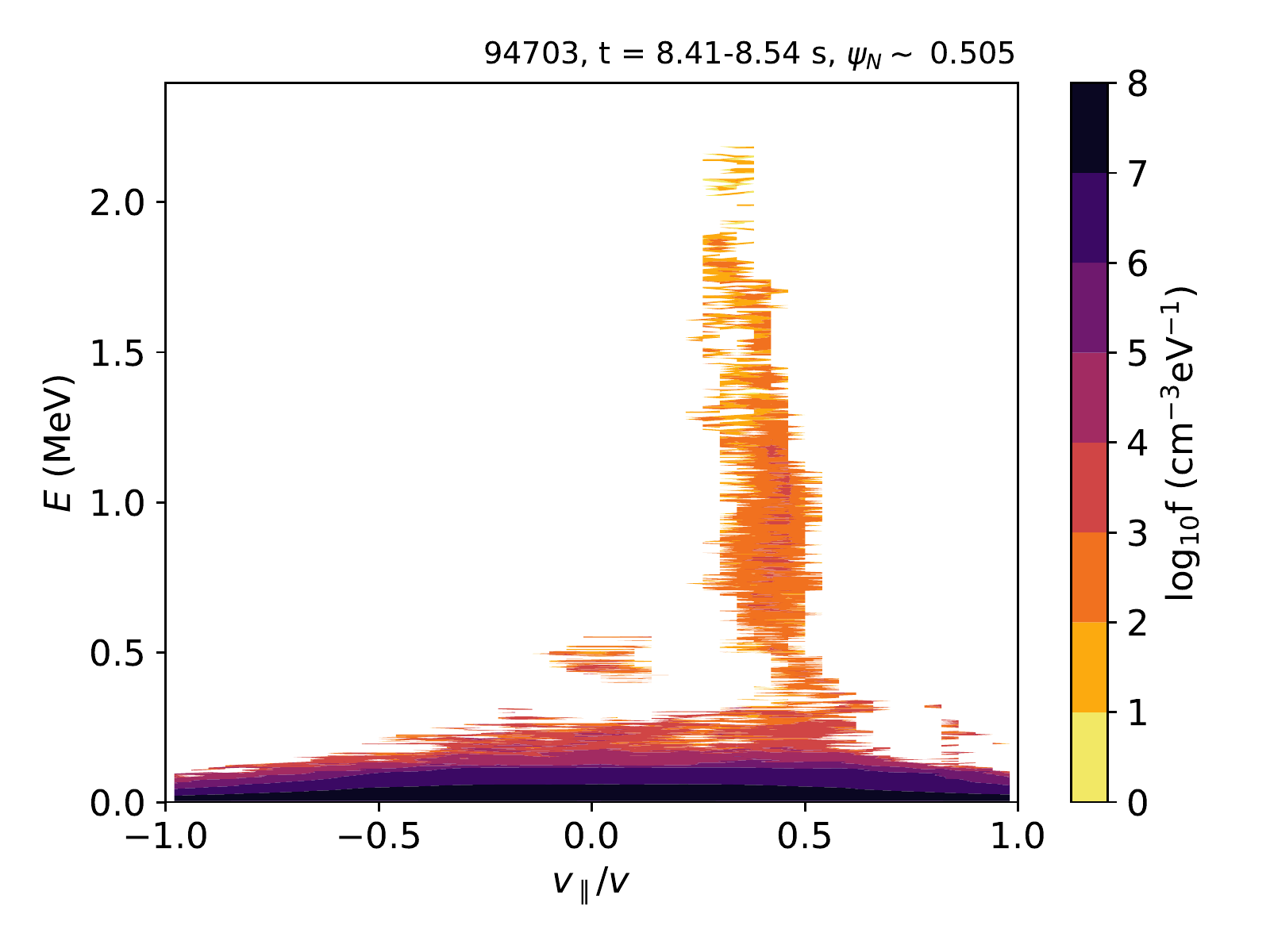}
            \caption{}
            \label{fig:fidf_pitchE}
        \end{subfigure}
        \caption{Deuterium fast ion distribution functions (logarithmic) from TRANSP for JPN~94703 and integrated over $\t = \SI{8.41{-}8.54}{s}$: (a)~averaged over all pitches, and (b)~at $\psin \approx 0.5$.}
        \label{fig:fidf_both}
    \end{figure}

    The pitch- and flux-surface-averaged \FI distribution is shown in \cref{fig:fidf_psiE}, while a ``slice'' at one flux surface ($\psin \approx 0.5$) is shown in \cref{fig:fidf_pitchE}. The bulk \NBI population is clearly seen below ${<}\SI{0.1}{MeV}$ with broad extent in radius and pitch. The \FI tail, accelerated by \ICRH, extends to ${\sim}\SI{2.5}{MeV}$, is primarily core-localized, and exhibits a dominant pitch $\vpar/v \approx 0.5$. In contrast to other pulses with the three-ion D-$\DNBI$-$\Hethree$ scheme at lower \NBI power \cite{Kazakov2021} with core-localized AEs, the \FI population did not destabilize \AEs in JPN~94703. As discussed next, the localization of the mode tracked by the \AEAD is such that drive from the \FI tail would not be expected; therefore, not including its effect on the growth rate is justified in this case.\footnotemark[\value{footnote}]
    
    
    \NOVAK simulations identify three candidate \AEs~- i.e. \EAEs in the edge gap - with similar frequencies and net damping rates as the experimentally measured mode. Their toroidal mode numbers, resonant frequencies, and breakdown of damping rate contributions are provided in \cref{tab:novak}. Note that the 23\% $\Hethree$ has not been simulated here,\footnotemark[\value{footnote}] 
    but its expected effect is to increase the \Alfven speed (and hence the plasma-frame \AE frequency) by only ${\sim}6\%$, which is approximately the same uncertainty introduced by any fitted profile (see \cref{fig:profiles}). \iaea{In addition, the ion dilution from including multiple thermal ion species would even further reduce the effect of ion Landau damping, discussed in more detail below.}
    
    \begin{table}[h!]
        \caption{Normalized damping rate (\%) calculated from \NOVAK. The frequency is in the lab frame, and uncertainties of continuum damping are $\pm 0.1\%$.}
        \label{tab:novak}
        \centering
        \begin{tabular}{l c c c}
            \hline
            Damping $\go$ (\%)  & $\n = 3, \fo = \SI{243.1}{kHz}$     & $\n = 5, \fo = \SI{236.4}{kHz}$   & $\n = 6, \fo = \SI{232.7}{kHz}$ \\
            \hline
            Continuum               & -0.092    & -0.116	& -0.177 \\
            Radiative               & 0.000     & 0.000	    & 0.000 \\
            Electron collisional    & -0.011    & -0.010    & -0.010 \\
            Electron Landau         & -0.161	& -0.198	& -0.176 \\
            Ion Landau 	            & $\sim$0.000	& $\sim$0.000	& $\sim$0.000 \\
            \NBI fast ions	        & -0.031	& -0.017	& -0.014 \\
            \textbf{Total}	        & \textbf{-0.295}	& \textbf{-0.341} 	& \textbf{-0.378} \\
            \hline
        \end{tabular}
    \end{table}

\begin{figure}[h!]
        \centering
        \begin{subfigure}{\halfwidth}
            \includegraphics[width=\textwidth]{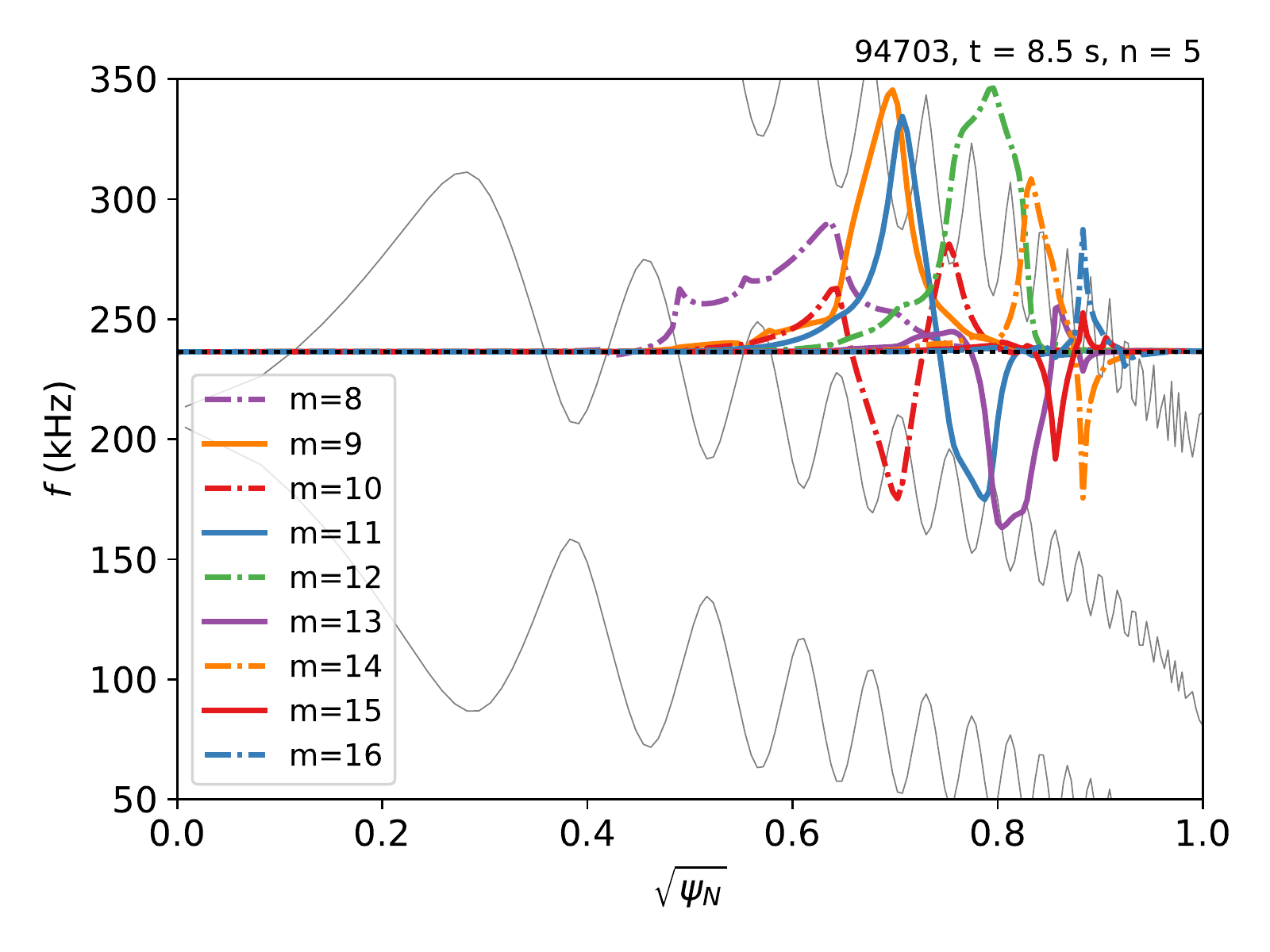}
            \caption{}
            \label{fig:novak5}
        \end{subfigure}
        \begin{subfigure}{\halfwidth}
            \includegraphics[width=\textwidth]{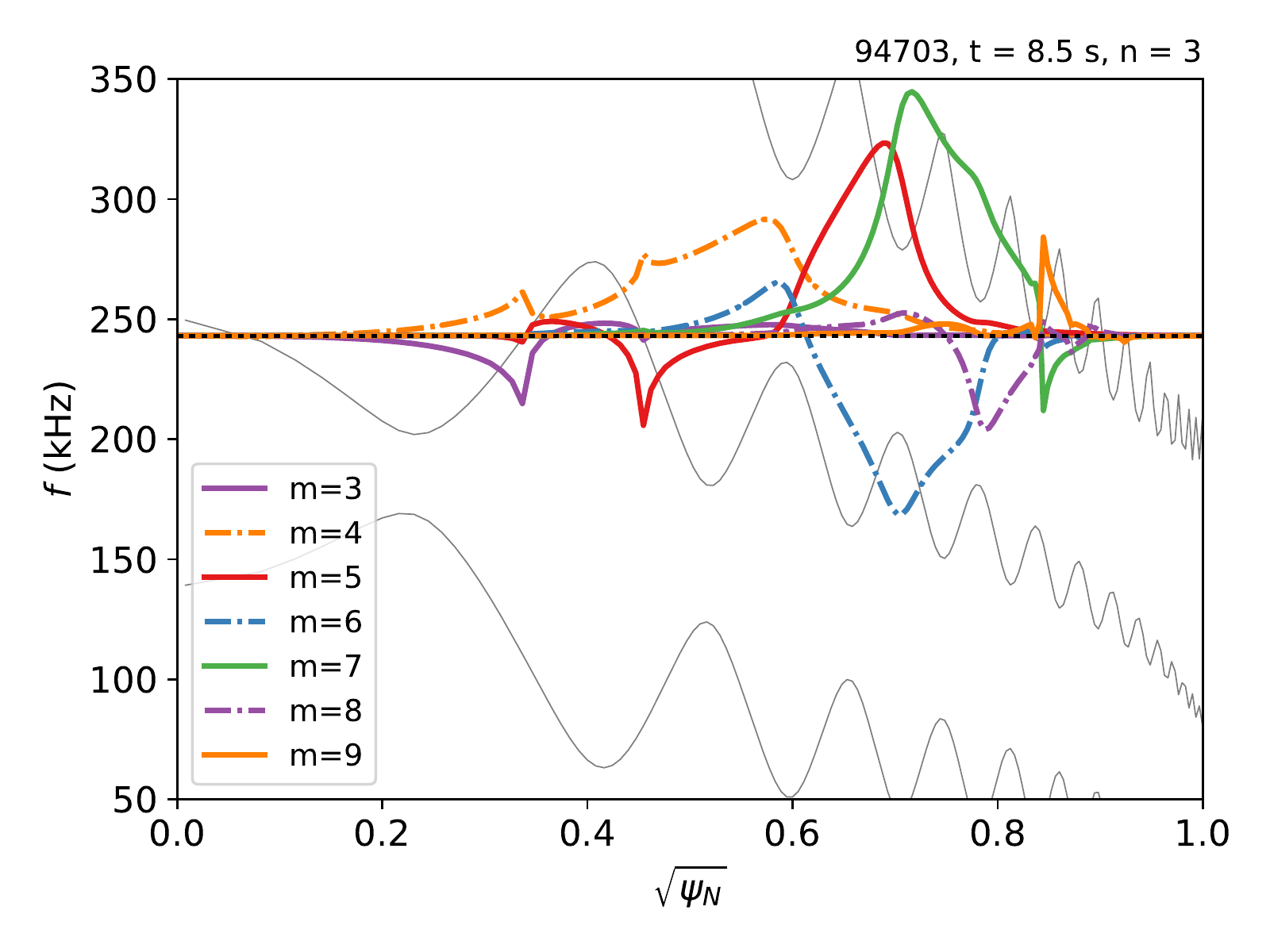}
            \caption{}
            \label{fig:novak3}
        \end{subfigure}
        \caption{Continua (thin lines) and poloidal mode structure (solid, dot-dashed) from \NOVAK at $\t = \SI{8.5}{s}$ for (a)~$\n = 5$ and (b)~$\n = 3$. The frequency (lab-frame) is indicated by the horizontal dotted line. $\psin$ is the normalized poloidal flux.}
        \label{fig:novak_both}
    \end{figure}

    The simulated $\n = 5$ mode, with resonant frequency $\fo \approx \SI{236}{kHz}$ and net damping rate $-\go \approx 0.34\%$, is the best match with the experimental \AE. The poloidal mode structure is plotted with the \Alfven continuum \yk{in the \emph{lab} frame} in \cref{fig:novak5}, with dominant couplings of poloidal harmonics $\m = (9,11)$, $(10,12)$, and $(11,13)$. Its localization around $\sqrt{\psin} \approx 0.7-0.9$ ($\psin \approx 0.5-0.8$) is consistent with improved \AEAD coupling to the edge as opposed to core modes, \iaea{which has also been observed in previous simulation work \cite{Borba2010,Dvornova2020,Tinguely2021}.}
    
    The simulated $\n=3$ mode is shown in \cref{fig:novak3} for comparison. Dominant poloidal harmonics are $\m = (5,7)$ and $(6,8)$. Note that the localization is similar to that of $\n=5$. Thus, there is certainly some ambiguity in toroidal mode number of the measured \AE. It could even be a superposition of modes. Nevertheless, \NOVAK is consistent in matching the frequency, net damping rate, and expected mode location.

    From \cref{tab:novak}, the dominant contributions to the \AE's marginal stability are electron Landau and continuum damping, with small contributions from \NBI fast ion and electron collisional damping. 
    \iaea{The explanation for the dominance of electron over ion Landau damping is likely due to the relatively high \Alfven speed, $\vA \approx \SI{7\times10^6}{m/s}$, compared to the electron and D ion thermal speeds, $\vte \approx 5.7\vA$ and $\vti \approx 0.1\vA$, respectively.}
    The intersections of the candidate modes and the \Alfven continua are clearly seen in \cref{fig:novak_both}.
    \iaea{The reason that the \AEAD's magnetic perturbation was able to ``tunnel through'' the edge continuum to excite this stable \AE may simply be due to the small magnitude of the continuum damping, only $\abs{\go} \approx 0.1\%{-}0.2\%$. This could help explain why only small damping rates are observed at high $\nnf$ in \cref{fig:n95}.}
    
    The almost negligible damping from \NBI fast ions makes sense since their parallel injection velocities, $\vpar \approx \SI{1.6\times10^6}{m/s}$, are less than \iaea{$\vA/2$, so there is little interaction with the candidate \EAEs}. Finally, note that the calculated radiative damping is totally negligible here. Thus, the damping mechanisms of this edge-localized \EAE are very different from those of some core-localized \TAEs in JET studied earlier with the \AEAD \cite{Aslanyan2019} \iaea{where radiative and ion Landau damping were dominant}. Yet this is expected given the different \AEs, localizations, and plasma scenarios.
    
    \yk{
    Finally, we note that the \Alfven continua for this pulse (see \cref{fig:novak_both}) exhibit an interesting feature: the \TAE and \EAE gaps exist at almost the same frequency but over different radial ranges - core and edge, respectively - due, in part, to the differential plasma rotation (see \cref{fig:profiles}) and associated Doppler shift. Though not shown here, this result is confirmed with the \Alfven continuum solver \CSCAS \cite{cscas_huysmans2001} using the magnetic geometry output from \HELENA \cite{helena_huysmans1991}. Such radially aligned gaps have been studied before - analytically and numerically - for reversed-shear $q$-profiles \cite{Gorelenkov2005}; however, in that study, core \EAE and edge \TAE gaps were aligned, \emph{opposite} to the present work.
    }
    
    \yk{In our case, it is reasonable to ask whether the \AEAD is resonating with a core \TAE instead of an edge \EAE.}%
    \footnote{\yk{
        Pursuing such \TAE solutions is not without some experimental motivation: there are plasma discharges very similar to JPN~94703 and also part of JET's three-ion D-$\DNBI$-$\Hethree$ heating scenario experiments (JPN~95683, for example \cite{Kazakov2021}), where core-localized \TAEs were driven by energetic ions. Further analyses of these plasmas should be pursued, but are beyond the scope of the present work.
    }}%
    \yk{
    A scan in the $\q$-profile - i.e. a translation up/down by $\pm0.1$ and $\pm0.2$ - in \NOVAK finds \emph{no} $\n=3$ or $5$ eigenmode solutions in the \TAE gap close to the experimentally measured frequency. Importantly, the $n=5$ \EAE is the most robust solution to this sensitivity study. However, by significantly lowering the central shear, an $\n=3$ \TAE solution appears near $\sqrt{\psin}\approx 0.4$ with frequency $\fo \approx \SI{235}{kHz}$. This is consistent with earlier observations that the three-ion \ICRH scenario can result in the flattening and even the reversal of the $q$-profile in the central regions of the plasma \cite{Kazakov2021}. However, since such profiles are outside of experimental uncertainties here, we conclude that our measurement corresponds to the edge-localized \EAE predicted by the most reasonable assumptions. 
    }

    \section{Summary}\label{sec:summary}
    
    Understanding the interaction of \AlfvenEigenmodes (\AEs) and energetic particles (\EPs) is important to the success of future tokamaks. In JET, eight toroidally spaced, in-vessel antennas - collectively called the \AEADiagnostic (\AEAD) - actively probe stable \AEs with frequencies ranging $f = \SI{25{-}250}{kHz}$ and toroidal mode numbers $\absn < 20$. The \AEAD plays an especially important role when \AEs are not destabilized by \EPs, which could even be the case for alpha drive in the upcoming JET DT campaign.

    During the 2019-2020 deuterium campaign, $\sim$7500 resonances – along with their frequencies $\wo = 2\pi\fo$, net damping rates $\g < 0$, and toroidal mode numbers $\n$ - were measured in $\sim$800 plasma discharges. A statistical analysis was performed on the database: continuum and radiative damping were found to increase with edge safety factor, edge magnetic shear, and when including non-ideal effects (see \cref{tab:correlations} and \cref{fig:lambda}), \iaea{as expected from theory}. A lower bound on ion Landau damping from \NBI fast ions was also inferred (see \cref{fig:nbi}).
    
    \iaea{
    By comparing the stable \AE database with the EUROfusion JET-ILW pedestal database, the \AEAD was determined to be inefficient in probing stable \AEs during \Hmode periods. One likely cause is reduced \AEAD accessibility due to interactions with the edge \Alfven continuum. This is consistent with stable \AE observations limited by high edge densities and steep edge density gradients (see \cref{fig:n95}). Reduced signal-to-noise is also a factor as ELMs can inhibit the identification of stable \AE resonances in the magnetics data. Finally, the number of stable \AE measurements and their corresponding damping rates were found to decrease with $\n$ (see \cref{fig:n}), possibly due to mode widths and damping becoming more localized.
    }
    


    A novel measurement was presented of a marginally stable \EAE resonantly excited by the \AEAD at the edge of an \Lmode/\Mmode plasma (JPN~94703) with high-power auxiliary heating, i.e. \ICRH and \NBI up to $\SI{25}{MW}$ (see \cref{fig:params_and_profiles,fig:fidf_both}). This stable \AE was tracked in real time with frequency $\fo \approx \SI{235{-}250}{kHz}$, net damping rate $-\go \approx 0.17\%{-}0.45\%$, and estimated toroidal mode number $\absn \approx 3{-}6$ (see \cref{fig:spectrogram_and_resonances}). 
    
    \NOVAK kinetic-MHD simulations showed good agreement with experimental measurements, indicating the dominance of electron Landau and continuum damping for a marginally stable edge-localized \EAE (see \cref{tab:novak,fig:novak_both}). \iaea{These dominant contributions, as well as negligible contributions from thermal ion and \NBI fast ion Landau damping, match physical intuition due to the relatively high \Alfven speed compared to thermal electron/ion and \NBI parallel injection velocities.} \yk{A scan of the $\q$ profile within experimental uncertainties indicated that the $n=5$ edge \EAE solution in JPN~94703 is robust, while a core/mid-radius \TAE solution is unlikely in this case.}
    
    With this demonstration of a successful \AE stability measurement by the \AEAD in a deuterium plasma with high-power auxiliary heating, we are optimistic for similar measurements in the ongoing tritium and upcoming DT campaigns.

    \section*{Acknowledgments}

    The authors thank 
    V.~Aslanyan,
    P.~Bonofiglo,
    N.~Dreval,
    L.~Frassinetti, 
    N.~Gorelenkov, 
    N.~Hawkes, 
    Ph.~Lauber,
    S.~Menmuir, 
    E.~Rachlew, 
    E.~Solano,
    G.~Szepesi,
    A.~Teplukhina, and
    D.~Testa
    for their contributions to this paper.
    \nfadd{The authors are also grateful to the reviewers who helped improve this paper.}
    This work was supported by US DOE grants DE-SC0014264 and DE-AC02-09CH11466,
    as well as the Brazilian agency FAPESP Project 2011/50773-0. This work has been carried out within the framework of the EUROfusion Consortium and has received funding from the Euratom research and training program 2014-2018 and 2019-2020 under grant agreement No 633053. The views and opinions expressed herein do not necessarily reflect those of the European Commission.

    \section*{References}
        \bibliographystyle{unsrt}

\end{document}